\documentclass[twocolumn,showpacs,amsmath,amssymb,floatfix,superscriptaddress]{revtex4}
\usepackage{graphicx}
\usepackage{dcolumn}
\usepackage{bm}
\usepackage{hyperref} 
\usepackage{latexsym}

\begin{document}
\title{Statistics of Lead Changes in Popularity-Driven Systems}
\author{P.~L.~Krapivsky}
\email{paulk@bu.edu}
\author{S.~Redner}
\email{redner@bu.edu}
\affiliation{Center for BioDynamics, Center for Polymer Studies, 
and Department of Physics, Boston University, Boston, MA, 02215}

\begin{abstract}
  
  We study statistical properties of the highest degree, or most popular,
  nodes in growing networks.  We show that the number of lead changes
  increases logarithmically with network size $N$, independent of the details
  of the growth mechanism.  The probability that the first node retains the
  lead approaches a finite constant for popularity-driven growth, and decays
  as $N^{-\phi}\,(\ln N)^{-1/2}$, with $\phi=0.08607\ldots$, for growth with
  no popularity bias.

\end{abstract}


\maketitle

Extremes are vitally important in science and engineering.  These quantities
are used to determine the likelihood of a rare event, such as the probability
of failure of a space shuttle launch or of a dam in flood conditions.  The
theory of extreme statistics provides a powerful tool to understand such
real-world situations \cite{Galambos}.  Extremes are also irresistible
in everyday life -- we are naturally drawn to compilations of various
pinnacles of human endeavor, such as, for example, lists of the most
beautiful people, the richest people, the most-cited scientists, athletic
records, {\it etc} \cite{guiness}.

This social perspective about extremes raises new questions for which much
less is known compared to the magnitude of the extreme value itself
\cite{MAA02}.  For example, how does the identity of the leader -- the
individual who possesses the extreme value of a particular attribute --
change as a function of time?  What is the rate at which lead changes occur?
What is the probability that a leader retains the lead as a function of time?

We address these questions within the framework of growing networks, where
the relevant quantity is the node degree -- the number of links that join to
each node.  We view the degree as quantifying the popularity (or wealth) of
the node, and the leader is the node with the highest degree.  We focus on
generic network models with preferential attachment to already popular nodes
\cite{yule,simon,BA,KR01,DMS2}, and networks with random attachment.  The
former describe, for example, the distributions of biological genera, word
frequencies, publications, urban populations, and income \cite{yule,simon},
and contemporary applications to collaboration networks and the World-Wide
Web have been developed \cite{applics}.  It has been posited that a hallmark
of such systems is ``the rich get richer'' -- that is, more popular nodes
tend to remain so \cite{simon,BA}.  Our basic goal is examine the
consequences of preferential and random attachment mechanisms in growing
networks and to test whether the adage of the rich get richer really does
apply.

The network grows by adding nodes, each of which links to a pre-existing node
with an attachment rate $A_k$ that depends only on the degree $k$ of the
target node.  We choose $A_k=k+\lambda$ with $\lambda>-1$ \cite{yule,simon}.
For such networks, the degree distribution has an asymptotic power-law tail,
$N_k\sim N/k^{3+\lambda}$, where $N_k$ is the number of nodes of degree $k$
and $N$ is the total number of nodes \cite{simon,KR01}.  For
$\lambda\to\infty$ the growth mechanism reduces to random attachment and the
degree distribution is exponential.

\begin{figure}[ht] 
 \vspace*{0.cm}
 \includegraphics*[width=0.45\textwidth]{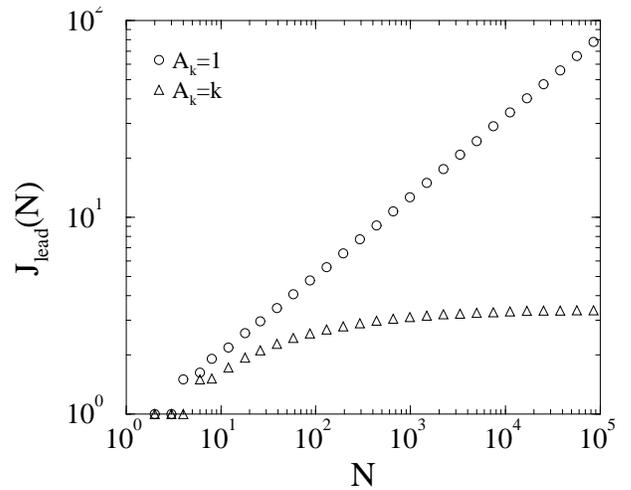}
\caption{Average index of the leader $J_{\rm lead}(N)$ as a
  function of the total number of nodes $N$ for $10^5$ realizations of a
  growing network of $10^5$ nodes.  Shown are the cases of attachment rates
  $A_k=1$ and $A_k=k$.}
\label{lead-index}
\end{figure}

{\it Identity of the leader}.  We characterize the identity of each node by
its index $J$.  A node of index $J$ is the $J^{\rm th}$ one introduced into
the network.  To start with an unambiguous leader, the initial system
contains $N=3$ nodes, with the initial leader having degree 2 (and index 1)
and the other two nodes having degree 1.  A new leader arises when its degree
exceeds that of the current leader.  For the linear attachment rate, $A_k=k$,
the average index of the leader $J_{\rm lead}(N)$ saturates to a finite value
of approximately 3.4 as $N\to\infty$ (Fig.~\ref{lead-index}).  With
probability $\approx 0.9$, the leader is from among the 10 earliest nodes,
while the probability that the leader is not among the 30 earliest nodes is
less than $0.01$.  Thus only the very earliest nodes have appreciable
probabilities to be the leader; the rich really do get richer.  Similarly in
the general case of $A_k=k+\lambda$, the average index of the leader also
saturates to a finite value that is a continuously increasing function of
$\lambda$.

For random attachment ($A_k=1$), the average index of the leader grows
algebraically, $J_{\rm lead}(N)\sim N^{\psi}$ with $\psi\approx 0.41$
(Fig.~\ref{lead-index}).  The leader is still an early node (since $\psi<1$),
but not necessarily one of the earliest.  For example, simulation indicates
that for $N=10^5$ a node with index greater than 100 has a probability of
approximately $10^{-2}$ of being the leader.  Thus, in random attachment, the
order of node creation plays a significant but not deterministic role in the
identity of the leader node.

The identity of the leader can be determined from the joint index-degree
distribution.  Let $C_k(J,N)$ be the average number of nodes of index $J$ and
degree $k$.  As shown in \cite{KR01}, for constant attachment rate, this
joint distribution has the Poisson form,
\begin{equation}
\label{ck0}
C_k(J,N)={J\over N}\,{|\ln(J/N)|^{k-1}\over (k-1)!}.
\end{equation} 
{}From this, the average index of a node of degree $k$ is
\begin{equation}
\label{av-J}
J_k(N) = {\sum_{1\leq J\leq N} J \,C_k(J,N)\over
        {\sum_{1\leq J\leq N} C_k(J,N)}} 
      = N\left({2\over3}\right)^k,
\end{equation}
implying $J_{\rm lead}(N)=N (2/3)^{k_{\rm max}}$.  We estimate the maximum
degree from the extreme value criterion $\sum_{k\geq k_{\rm max}}
N_k(N)\approx 1$.  Using $N_k(N)= N/2^k$ \cite{KR01}, we find $2^{k_{\rm
    max}}\approx N$, or $k_{\rm max}\sim{\ln N/\ln 2}$.  Therefore
\begin{eqnarray*}
J_{\rm lead}(N)\propto N^\psi,\quad {\rm with}\quad
\psi=2-{\ln 3\over \ln 2} \approx 0.415\,037,
\end{eqnarray*}
in excellent agreement with our numerical results.

For the linear attachment rate, the joint index-degree distribution
is \cite{KR01}
\begin{equation}
\label{ck1all}
C_k(J,N)=\sqrt{J\over N}\left\{1-\sqrt{J\over N}\right\}^{k-1},
\end{equation}
{}from which the average index of a node of degree $k$ is $J_k(N) =
12N/[(k+3)(k+4)]$.  Since $N_k(N) \sim 4N/k^3$ for the linear attachment rate
\cite{BA,KR01}, the extreme statistics criterion $\sum_{k\geq k_{\rm max}}
N_k(N)\approx 1$ gives $k_{\rm max}\approx \sqrt{N}$.  Therefore $J_{\rm
  lead}(N)\sim 12N/ k_{\rm max}^2={\cal O}(1)$ indeed saturates to a finite
value.  A similar result holds in the general case $A_k=k+\lambda$.  Thus the
leader is one of the first few nodes in the network.

\begin{figure}[ht] 
  \vspace*{0.cm} \includegraphics*[width=0.45\textwidth]{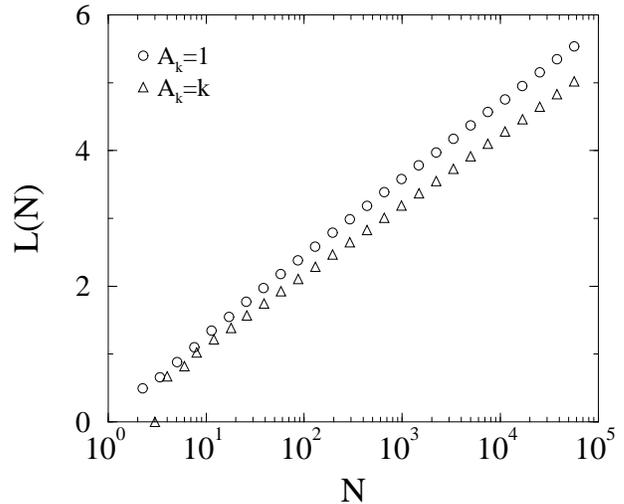}
\caption{Average number of lead changes $L(N)$ as a function of network 
  size $N$ for $10^5$ realizations of the network for $A_k=1$ and $A_k=k$.}
\label{lead}
\end{figure}

{\it Lead changes}.  We find that the average number of lead changes $L(N)$
grows logarithmically in $N$ for both the attachment rates $A_k=1$ and
$A_k=k$ (Fig.~\ref{lead}).  There is, however, a significant difference in
the distribution of the number of lead changes, $P(L)$, at fixed $N$.  For
$A_k=1$, this distribution is sharply localized, with the average value
$L\approx 5.609$ in a network of $N=10^5$ nodes, while the maximum number of
lead changes in $10^5$ realizations was 16.  On the other hand, for $A_k=k$,
$P(L)$ has a significant tail and the maximum number of lead changes is 63.
This longer tail in $P(L)$ for linear attachment stems from repeated lead
changes among the two leading nodes.  Even though the distribution is
visually broader, the average number of lead changes, $L\approx 5.096$, is
less than that for $A_k=1$.  Related to lead changes is the number of {\em
  distinct\/} nodes that enjoy the lead over the history of the network.
Simulations indicate that this quantity also grows logarithmically in $N$.

This logarithmic behavior can be easily understood for the attachment rate
$A_k=1$.  Here the number of lead changes cannot exceed the upper bound given
by the maximal degree $k_{\rm max}\sim \ln N/\ln 2$.  To establish the
logarithmic growth in the general case we first note that when a new node is
added, the lead changes if the leadership is currently shared between two (or
more) nodes and the new node attaches to a co-leader.  The number of
co-leader nodes (with degree $k=k_{\rm max}$) is $N/k_{\rm max}^{3+\lambda}$,
while the probability of attaching to a co-leader is $k_{\rm max}/N$.  Thus
the average number of lead changes satisfies
\begin{equation}
\label{LN}
{d\over dN}\,L(N)\propto 
{k_{\rm max}\over N}\,{N\over k_{\rm max}^{3+\lambda}}.
\end{equation}
Since the maximal degree $k_{\rm max}$ grows as $N^{1/(2+\lambda)}$,
Eq.~(\ref{LN}) reduces to ${d L(N)/dN}\propto N^{-1}$ and thus gives the
logarithmic growth $L(N)\propto \ln N$.  This argument can be adapted to
networks with arbitrary attachment rates (except those growing faster than
linearly with $k$ \cite{KR01}), and thus the growth law $L(N)\propto \ln N$ is
universal.  This universality is reminiscent of the radius of random networks
which typically are proportional to $\ln N$, independent of their
construction mechanism.

\begin{figure}[ht] 
 \vspace*{0.cm}
 \includegraphics*[width=0.45\textwidth]{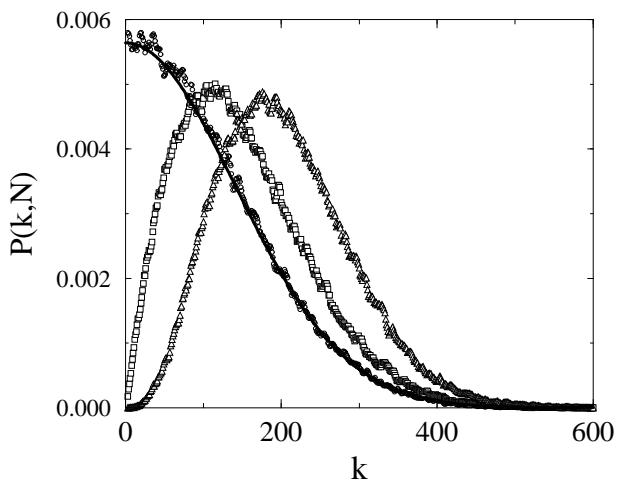}
\caption{Degree distribution of the first node for the dimer, trimer
  (2 links to the initial node), and pentamer (4 links) initial conditions
  based on $10^5$ realizations of a network with $10^5$ links.  The curve is
  the prediction of Eq.~(\ref{P-final}).}
\label{P1k}
\end{figure}

{\it Fate of the first leader}.  Figure \ref{P1k} shows that the degree
distribution of the first node depends on the initial conditions for the
linear attachment rate; the same is true in the general case $A_k=k+\lambda$
while for $A_k=1$ the initial condition is asymptotically irrelevant. 

We can determine the degree distribution of the first node analytically for
the constant and linear attachment rates.  (A similar approach is given in
Ref.~\cite{DMS2}).  Let $P(k,N)$ be the probability that the first node has
degree $k$ in a network of $N$ links \cite{note2}.  For $A_k=k$, this
probability obeys the master equation
\begin{equation}
\label{P}
P(k,N+1)= {k-1\over 2N}\, P(k-1,N)+{2N-k\over 2N}\,\,P(k,N).
\end{equation}
The first term on the right accounts for the situation when the earliest node
has degree $k-1$.  Then a new node attaches to it with probability
$(k-1)/2N$, thereby increasing the probability for the node to have degree
$k$.  Conversely, with probability $(2N-k)/2N$ a new node does not attach to
the earliest node, thereby giving the second contribution to $P(k,N+1)$.

The solution to the master equation (\ref{P}) for the ``dimer'' initial
condition $\circ$\!\rule[.03in]{.2in}{.01in}\!$\circ$ is \cite{KR02}
\begin{equation}
\label{Pdimer} 
P(k,N)={1\over 2^{2N-k-1}}\, {(2N-k-1)!\over (N-k)!\,(N-1)!}\,.
\end{equation}
For $N\to\infty$, this simplifies to the Gaussian distribution
\begin{equation}
\label{P-final}
P(k,N)\sim {1\over\sqrt{\pi N}}\,\,e^{-k^2/4N}
\end{equation}
for finite values of the scaling variable $k/N^{1/2}$.  Thus the typical
degree of the first node is of the order of $N^{1/2}$; this is the same
scaling behavior as the degree of the leader node.  For the trimer initial
condition (which we typically used in simulations) we obtained the degree
distribution of the first node in the form of a series of ratios of gamma
functions \cite{KR02}, in which $P(k,N)$ has an $e^{-k^2/4N}$ Gaussian tail,
independent of the initial condition.  The degree of the first node also
approximates that of the leader node \cite{MAA02} more and more closely as
the degree of the first node in the initial state is increased.

Although $P(k,N)$ contains all information about the degree of the first
node, the behavior of its moments $\langle k^a\rangle_N=\sum k^a P(k,N)$ is
simpler to appreciate.  To determine the moments, it is more convenient to
construct their governing recursion relations directly, rather than to
calculate the moments from $P(k,N)$.  Using Eq.~(\ref{P}), the average degree
satisfies the recursion relation
\begin{equation}
\label{k-rec}
\langle k\rangle_{N+1}=\langle k\rangle_{N}\left(1+{1\over 2N}\right)\,,
\end{equation}
whose solution is
\begin{equation}
\label{k-soln}
\langle k\rangle_N=\Lambda\,{\Gamma\left(N+\frac{1}{2}\right)\over
{\Gamma\left(\frac{1}{2}\right)\Gamma(N)}}\sim 
{\Lambda\over\sqrt{\pi}}\, N^{1/2}\,.
\end{equation}
The prefactor $\Lambda$ depends on the initial condition, with $\Lambda=2,
8/3, 16/5,\ldots$ for the dimer, trimer, tetramer, {\it etc.}, initial
conditions.

This multiplicative dependence on the initial condition means that the first
few growth steps substantially affect the average degree of the first node.
For example, for the dimer initial condition, the average degree of the first
node is, asymptotically, $\langle k\rangle_N\sim 2\sqrt{N/\pi}$.  However, if
the second link attaches to the first node, an effective trimer initial
condition arises and $\langle k\rangle_N\sim (8/3)\sqrt{N/\pi}$.  Thus small
initial perturbations lead to huge differences in the degree of the first
node.

\begin{figure}[ht] 
 \vspace*{0.cm}
 \includegraphics*[width=0.45\textwidth]{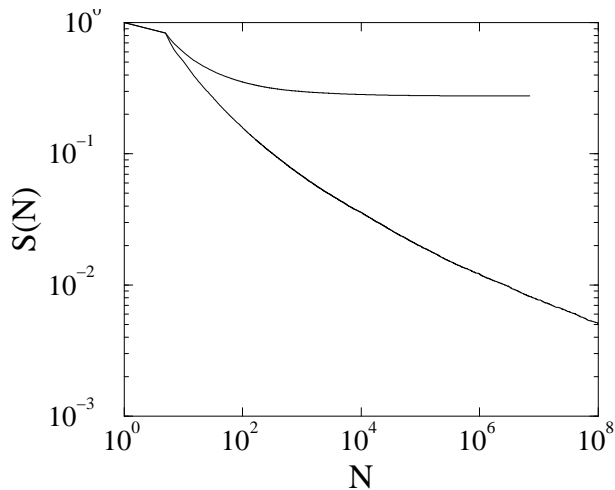}
\caption{Probability that the first node leads throughout the evolution 
  for $10^5$ realizations of up to size $N=10^7$ for $A_k=k$ (upper), and up
  to $N=10^8$ for $A_k=1$ (lower).}
\label{surv}
\end{figure}

An intriguing manifestation of the rich get richer phenomenon is the behavior
of the survival probability $S(N)$ that the first node leads throughout the
growth up to size $N$ (Fig.~\ref{surv}).  For the linear attachment rate,
$S(N)$ saturates to a finite non-zero value of approximately 0.277
as $N\to\infty$; saturation also occurs for the general attachment rate
$A_k=k+\lambda$.  Thus for these popularity-driven systems, the rich get
richer holds in a strong form -- the lead never changes with a positive
probability.  

For constant attachment rate, $S(N)$ decays to zero as $N\to\infty$, but
asymptotic behavior is not apparent even when $N=10^8$.  A power law
$S(N)\propto N^{-\phi}$ is a reasonable fit, but the local exponent is still
slowly decreasing at $N\approx 10^8$ where it has reached $\phi(N)\approx
0.18$.  To understand the slow approach to asymptotic behavior, we study the
degree distribution of the first node.  This quantity satisfies the recursion
relation
\begin{equation}
\label{P0}
P(k,N)={1\over N}P(k-1,N-1)+{N-1\over N}P(k,N-1)
\end{equation}
which reduces to the convection-diffusion equation
\begin{equation}
\label{Pcont}
\left({\partial \over\partial \ln N}+{\partial \over\partial k}\right)
P={1\over 2}\,{\partial^2 P\over\partial k^2}
\end{equation}
in the continuum limit.  The solution is a Gaussian
\begin{equation}
\label{Pgauss}
P(k,N)={1\over \sqrt{2\pi \ln N}}\,
\exp\left\{-{(k-\ln N)^2\over 2\ln N}\right\}.
\end{equation}
Therefore the degree of the first node grows as $\ln N$, with fluctuations of
the order of $\sqrt{\ln N}$.  On the other hand, the maximal degree grows
faster, as $v \ln N$ with $v=1/\ln 2$, and its fluctuations are negligible.

We now estimate the large-$N$ behavior of $S(N)$ as $\sum_{k\geq k_{\rm
    max}}P(k,N)$.  This approximation gives
\begin{eqnarray}
\label{phi}
S(N)&\propto& \int_{v\ln N}^\infty {dk\over \sqrt{\ln N}}
\exp\left\{-{(k-\ln N)^2\over 2\ln N}\right\}\nonumber\\\nonumber\\
&\propto&N^{-\phi} \,\,(\ln N)^{-1/2}\, ,
\end{eqnarray}
with $\phi=(v-1)^2/2\approx 0.097989\ldots$.  The logarithmic factor leads to
a very slow approach to asymptotic behavior.

The above estimate is based on the Gaussian approximate for $P(k,N)$ which is
not accurate outside the scaling region, namely, for $k\gg\ln N+ \sqrt{\ln
  N}$.  However, we can determine $P(k,N)$ exactly because its defining
recursion formula, Eq.~(\ref{P0}), is closely related to that of the Stirling
numbers of the first kind ${N\brack k}$ \cite{stir}, and the solution for the
dimer initial condition is $P(k,N)={N\brack k}/N!$.  The corresponding
generating function is
\begin{eqnarray*}
S_N(x)=\sum_{k=1}^N P(k,N)\,x^k={x(x+1)\ldots(x+N-1)\over N!}\,.
\end{eqnarray*}
Using the Cauchy theorem, we express $P(k,N)$ in terms of the contour
integral $S_N(x)/x^{k+1}$.  When $N\to\infty$, this contour integral is
easily computed by applying the saddle point technique \cite{KR02}.  Finally
we arrive at Eq.~(\ref{phi}) with the same logarithmic prefactor but with a
slightly smaller {\em exact} exponent $\phi=1-v+v\ln v\approx 0.08607$.

In summary, lead changes are rare in popularity-driven network growth
processes and leadership is restricted to the earliest nodes.  With finite
probability, the first node remains the leader throughout the evolution.  For
growth with no popularity bias, leadership is shared among a somewhat larger
cadre of nodes.  As a consequence the average index of the leader node grows
as $N^\psi$ with $\psi= 0.415\,037\ldots$.  The possibility of sharing the
lead among a larger subset of nodes gives a rich dynamics in which the
probability that the first node retains the lead decays as $N^{-\phi}
\,\,(\ln N)^{-1/2}$ with $\phi=0.08067\ldots$.

We are grateful to NSF grant DMR9978902 for partial financial support of this
research.


\begin{thebibliography}{99}
\bibitem{note2} The normalized attachment probability is $A_k/A$, with
  $A=\sum A_j N_j$.  For the linear attachment rate, $A$ is twice the total
  number of links.  Hence formulae become more neat if we denote by $N$ the
  total number of links.
\end{thebibliography}
\end{document}